# DECIPHERING SOCIAL OPINION POLARIZATION TOWARDS POLITICAL EVENT USING TOPIC MODELLING AND DYNAMIC NETWORK ANALYSIS

Andry Alamsyah, Wachda Yuniar Rochmah, Arina Nahya Nurnafia

School of Economics and Business, Telkom University, Bandung, Indonesia

andrya@telkomuniversity.ac.id

*Abstract--* We have seen that some of social opinion polarization leads to the breakup of the relationship, some in the scale of small communities, but others can divide large organizations or even a nation. The legacy methodology to answer the root cause of opinion polarization in the society commonly use random sampling and questionnaire approach. This approach generally expensive in term of time and money. On the contrary, we have the opportunity to employ big data approach using social media data. Big data methodology provides us the rich source to investigate several questions, such as: how social opinion polarization formed, dynamic social network mechanism during time-windows observation, identification of dominant actors and communities. The power of big data approach lies in the number of data analyzed, where the more data involved in the process, the more accurate to describe the population condition. Today, computing power is no longer become an obstacle to process large volume, fast, and variety data, thus the observation of social opinion polarization process is possible.

In this research, we answer the social opinion polarization root cause by using topic modelling methodology. The dynamic social network mechanism is measured using social network properties. Identification of influential actors and communities are provided by social network analysis metric and methodology. By answering three major questions above, we are able to explain of opinion polarization and its growth along the time, both for their topology structure and conversational content. The process knowledge gives us insight to how and when the separation takes place.

As a case study, we use two massive adverse political campaign in Indonesia regarding the opinion of pro and contra the incumbent president to continue his presidentship in 2019 presidential election. We benefited of Indonesian habit to produce user-generated content in term of post and conversation in social media. We collect Twitter data with the following information: the observation duration is 10 days from April 27$^{th}$ to May 2$^{nd}$, 2018, data collected based on several opposite hashtags such as *#2019GantiPresiden #Jokowi2Periode* and many more, we acquire 24097 and 418256 tweets from pro and contra movement. The benefit of this study is to avoid the danger of deepening gap between opponent communities by implementing right strategy to prevent that from happen.

## I. INTRODUCTION

Technology has permitted us to communicate efficiently, thus provide medium for societies to growth. Human interaction brings knowledge, influencing opinion, disseminate information, and grouping people of similar interest or opinion. This online activity leaves digital trace behind, which benefit us to capture human behavior. Data analytics provides set of tools and methodology that allows us to extract knowledge regarding specific needs related to social problem. Large-scale data generated from social interactions lead to the formation of new science of society, called network science. The opportunity to form a new high-quality level using data has been widely used from social media.

The social media contents, whether in the form of images, videos, testimonials, tweets, blog posts, reviews and many more are considered as User Generated Content (UGC). Many people participate in UGC creation as part of their effort to build social reputation, or just finding the right information. UGC is a substantial resource as it denotes many precious insights whom social scientists seek. We could employ data analytics methods to find the pattern from UGC data, which mainly is in the form of unstructured data. The nature of unstructured data increases the processing complexity and in many cases the data cannot be processed at all using legacy procedure such as data mining [1]. One of the fastest strategies to process the unstructured data from social media is by using Social Network Analysis (SNA) methodology. SNA model actors and their surrounding relationship with their neighbors' actors. To capture the dynamic change of growth and shrink relationship between actors, we use Dynamic Network Analysis (DNA) [2].

There are many network measurements available to use to quantify a social network. This is also called as network properties. We are able to track the evolution of network grace to these metrics. The network metrics are as follows: nodes represents actors, edges represent relationship among actors, network average degree determine the number of relationships each actor has divided by the total number of relationships occur in overall network, network diameter shows the maximum distance from each actors in network ends, modularity to measure the tendency of network to clustered, and network density measure the ratio of current edges in the network to maximum number edges possible [3].

One of most dynamic social conversation is political domain. Towards an election event, society will tend to support a political figure according to their preferences. Some consider the vision and mission, track record, achievement, or even political party background. Eventually, these preferences will lead to opinion polarization in our society. With the development of technology, it greatly facilitates the public express their opinion in social media. This phenomenon gives us advantage to see the tendency that occurs in the community, either in favor or against a political figure.

Jokowi is $7^{th}$ president of Indonesia who will join the president election in 2019. A large number of people support him to run for president for the second time. Nonetheless, there are people who don't. These people who oppose Jokowi express their disapproval through a campaign called *Ganti Presiden*. To know public opinion on this matter, we extract some information from *Twitter* as a popular social media platform in Indonesia. Our datasets contain tweets by different hashtags. We separate the sentiment through the hashtag we use, which may lead to pro and contra to Jokowi. In order to understand what is being discussed by certain communities on Twitter, we utilize a method called Topic Modeling. We are able to discover the topics being discussed through the most probable terms within topics. Finally, we use SNA and DNA methodology to measure network metrics created by pro and contra hashtags.

User interaction in social media form a social network in some way similar to real-world networks as it represents graph with set of nodes as actors and edges as actor's relationship [3]. SNA is a way to quantify various interaction pattern in social network. It is capable to feed researcher understanding the ensemble of actors' characteristics, most influential actors, detect network community and several other measurements [4].

DNA is an emergent scientific field that cumulate traditional SNA within network science and network theory. There are two aspects of this field. The first one is the statistical analysis of DNA data. The second is the utilization of simulation to address issues of network dynamics. Differences between DNA networks from traditional social networks are larger scale, dynamic, complex networks, and may contain varying levels of uncertainty [5]. DNA also takes interactions of social features conditioning structure and behavior of networks into record. The evolution of dynamics actor interactions gives valuable insights about actors' online social behavior. This DNA permits us to comprehend how relationships thrive from time to time, relationship created among actors and how information diffuses [6].

In this paper, we investigate the social opinion polarization mechanism using Topic Modelling on each opponent's side split by Topic Modelling (TM). TM help us identify dominant terms on each opponent's side. It is followed by Text Network Analysis (TNA) methodology to summarize conversation on each side. And at last the DNA mechanism is measured using social network properties. Identification of influential actors and communities are provided by SNA metric and methodology. By answering three major questions: how social opinion polarization formed, dynamic social network mechanism during time-windows observation, identification of dominant actors and communities, we are able to explain of opinion polarization and its growth along the time, both for their topology structure and conversational content. The process knowledge gives us insight to how and when the separation takes place.

## II. LITERATURE REVIEWS

### A. Topic Modeling (TM)

Topic Modeling (TM) is an approach to infer the topic in a document. The algorithm we use is Latent Dirichlet Allocation (LDA), which is a generative probabilistic model of a corpus. Documents are represented as random mixtures over latent topics, where each topic is indicated by a distribution over words [7].

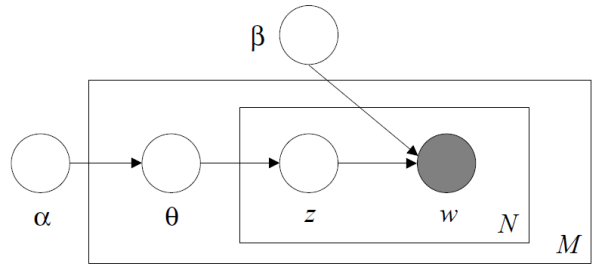

*Fig.1 The Latent Dirichlet Allocation model illustration*

The LDA model is portrayed as a probabilistic graphical model in Figure 1. The boxes are plates that represents replicates. The outer plate represents documents, while the inner plate represents the repeated choice of topics and words within a document. The parameters α and β are parameters in the corpus-level, assumed to be sampled once in the corpus generating process. The variables $θ_d$ are variables in the document-level, sampled once per document. At last, the variables $Z_{dn}$ and $W_{dn}$ are word-level variables and are sampled once for each word in each document [7].

B. **Social Network Analysis (SNA)**

Social Network Analysis (SNA) is an exploration and describing patterns approach in social relationship formed by an individual or group. Interactions occur in social network can be represented into two elements: set of entity that represents actors or individuals called nodes and sets of entity that represents relationship or interaction occur among actors or groups within the whole network called edges. SNA has become an important research that focus in many areas, such as: management, sociology, health care and many more. People interact with each other online, they barely rely on paper-based questionnaires results to build network, mainly due to the limitation of data acquisition. SNA has four main basic concepts: how to model actors' interactions into social networks, how to represent tie strength relationships between actors, how to identify key or important actors in the networks, and how to measure the structure of network cohesion [2].

C. **Dynamic Network Analysis (DNA)**

Dynamic Network Analysis (DNA) consists of analytic and algorithmic models that identify the overall process of social network evolution to forecast individual as well as group behavior and their relationships to each other. DNA is the latest approach that can figure interaction and network analysis up to the more complex thinking. DNA captures the dynamics network structures through complex systems as sequences of time within interactions [6] [8].

Dynamic Network properties are formed by several stages. Edge evolution shows the dynamics changes in interaction or relationships that occur within period of time. Node evolution shows the dynamic changes in the number of nodes or actors exist from time to time. Diameter evolution shows the dynamics of network size formed over time in order to see the value and the density. Last but not least, average degree evolution shows the growth of the average number of interactions each node makes within a certain period of time.

D. **Text Network Analysis**

Text Network Analysis (TNA) is one of the possible ways to represents the text occurrence complexity. The nodes consist of the texts and the relationship of the texts are represented by the edge [9]. A way to summarize a document is by represent it as a network of document text. This network contains the nodes or the texts and the edges or the relations of the texts which represent text occurrences on the same phrase. Higher frequency of texts occurrence next to each other in a phrase results stronger connection between them. The meanings and agendas that can arise from that interconnectedness are very diverse. However, each expression of the text has a certain purpose, related to a certain moment of time. Having the text as a network allows for a much more holistic views of the text and for many other expressions of the same agenda that could be more related to a specific context. [10].

## III. METHODOLOGY

Our methodology is started with data collection process, followed by data preprocessing, main process, and at last summarization of overall process. The methodology of our research framework shown in Figure 2. The details of each process explained in sub chapter lll A, B, C, and D.

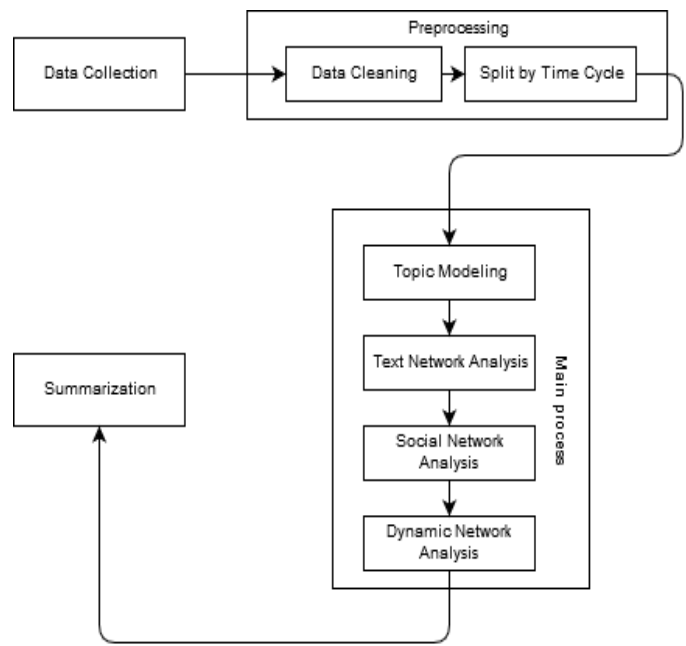

*Fig.2 Research Workflow*

### A. Data Collection Process

At the first stage, we collect *Twitter* data stream from April 27th until May 2nd, 2018 filtered by pro and contra hashtags. We use *Twitter* Application Programming Interface (API) as the gate to access public *Twitter* data. The pro and contra hashtag used as the filter shown in Table 1. Table 1 also shown the number of tweets collected on each side.

*Table 1 The Hashtag List*

|  | **Pro** | **Contra** |
|---|---|---|
| **Hashtag** | *jokowi2periode, JKW2P, jokowipresiden2019, 2019tetapjokowi, jokowisekalilagi, rakyatmaujokowi2019, jokowiduaperiode, salam2jari, ogah2019gantipresiden, diasibukkerja.* | *2019gantipresiden, 2019presidenbaru, gantipresidenyuk, gantipresiden, gantipresiden2019, asalbukanjkw, 2019gantirezim, 2019wajibgantipresiden, 2019asalbukanjokowi.* |
| **Number of Tweets** | 24097 | 418256 |

### B. Data Preprocessing

1) Data Cleaning

Data cleaning is the first sub process, which is crucial stage to reduce the complexity of quantification process leading to better input to the next process. Basically, in this stage, for TM and TNA, we remove unnecessary words through process such as stopwords removal, tokenization to separate phrase to words, stemming to transform the word into original form. For SNA and DNA, we clean the datasets from all those bots, users who randomly tend to spam by selling things online, and all tweets without any interaction in order to filter the source and target actor's interaction automatically. The table 2 gives the examples of the cleaning process for the TM and TNA.

*Table 2 The Example of Data Cleaning*

| Before | After |
|---|---|
| *Mengintimidasi* | *intimidasi* |
| *Memilih* | *pilih* |
| *Kaus* | *kaos* |

2) Split by Time Cycle

To see the evolution of network properties within the overall network from both sides, we split the datasets day by day. By splitting the dataset, we are able to see the growth or shrink the topology within the overall network from ten days observation time. This preprocessing step is done merely for DNA process.

**C. Main Process**

We have 4 main processes which consist orderly as follows: TM, TNA, SNA, and DNA. Each process has specific function to support the end summarization in research conclusions. TM is used to detect the topic based on each pro and contra hashtag tweets. SNA produces network model and its properties from of each pro and contra network. Meanwhile, DNA is used to track network evolution during observation time. The last process is TNA, which is basically same as SNA, but we use texts rather than actors. The explanation of each part of main process is as follows:

1) Topic Modeling (TM)

LDA methodology to detect topics is implemented on each opponent's side. LDA generate topics based on word frequency from each side tweets data. Here that means 24097 pro tweets and 418256 contra tweets. LDA is reasonably fast and accurate to find mixtures topic in collection of documents. From each opponent's side, we map the topics generated and see the terms arranging each topic based on frequency of appearance. Hence, we may see what the topic emerges from those collection of terms.

2) Social Network Analysis (SNA)

The social network construction is based from actor's interaction tweet. Once we have the tweet data, we transform raw tweets into node source and node target list represent them as edge list. We only collect user name who tweet and user name who is mentioned in the particular tweet. The mention mechanism could be in the form of conversations or retweets. If someone tweet without generate the conversations, thus we can omit those particular tweets. Figure 3 show (a) the raw tweet format and (b) the list of node source and target. Edge list in Figure 3 is the main ingredients to construct social network related to political and the election event.

(a)     (b)

*Fig.3 (a) Raw Tweet Format. (b) Node Source and Target List*

3) Dynamic Network Analysis (DNA)

DNA exerted to identify the dynamics interaction among actors in tweet universe. DNA capture each SNA properties measurement on different observation time. In this research, our measurement based on daily observation. As the result, we have graph or network evolution over time. We illustrate the data by the chart, where x-axis represents the period of time and y-axis represents the network properties value. This visualization enables us to see network and actor's behavior of over time.

4) Text Network Analysis (TNA)

TNA is basically work like SNA model, where nodes represent texts or terms instead of actors. Edge in TNA represents cooccurrence texts in the same phrase or document. This method helps us to summarize the large-scale document or tweet content into condense network information. Compared to the common method of extracting information from document such as word cloud, TNA have the advantage of having easier interpretation, since the present of link between texts show how relatable those texts in the document or tweet. We apply TNA for each pro and contra side by the following work order: 1. finding the cooccurrence between texts in the same tweet, 2. measure each text frequency of appearance, 3. measure the intensity of relations, 4. detect network modularity to see grouping texts between different topics, 5. construct the network, where node size represent frequency of appearance, edge size represent intensity of relations, and node – edge color as the text group modularity.

## IV. RESULT AND ANALYSIS

**A. Topic Modelling (TM)**

**1) Pro Hashtags**

We obtain the top 5 biggest topics for each pro and contra dataset. In the figure.3, we see the distribution of topics in the of pro hashtag. Each topic consists of several domain text or terms. Afterward, the figure 4,5,6,7 show the distributions of terms in each topic respectfully ordered by the biggest to the lowest. The blue colored bar represents the frequency of overall terms in the topics, meanwhile, the red colored bar shows the frequency of term in the current topic.

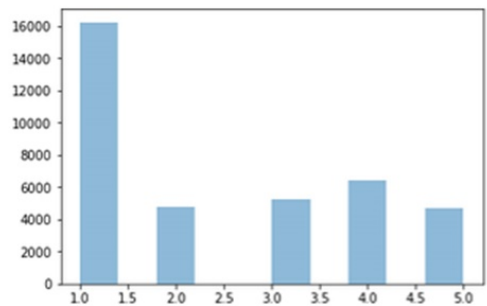

```
In total there are 5 major topics, distributed as follows
```

```
Printing top 7 Topics, with top 7 Words:
Topic #0:
negara kaos hoax indonesia nkri politik gerindra
Topic #1:
presiden prabowo ganti jokowimembangunindonesia anak cfd bandung
Topic #2:
diasibukkerja orang masyarakat hastag buruhtetapjokowi muslim pendukung
Topic #3:
jokowi indonesia rakyat dukung jokowisekalilagi joko widodo
Topic #4:
kerja jokowisekalilagi foto tagar tka jokowipresidenku kaum
```

*Fig.3 Top 5 biggest topics of pro hashtags*

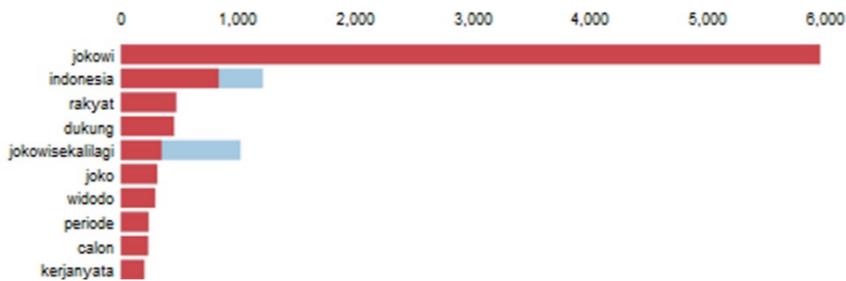

*Fig.4 First topic of pro hashtags*

The first topic in the pro dataset in figure 4. shows the public support towards incumbent president Joko Widodo or Jokowi to step forward for the next period. We take the above conclusion by the presence of these terms: *Jokowi, dukung, jokowisekalilagi, periode,* and *kerjanyata*. *Jokowi* is the most dominant term showing up, which is up to 6000 times. Public mostly talk about positive moves that Jokowi has created during his tenure as president, including his strength, his spirit, and his determination to implement transparency and fight corruption.

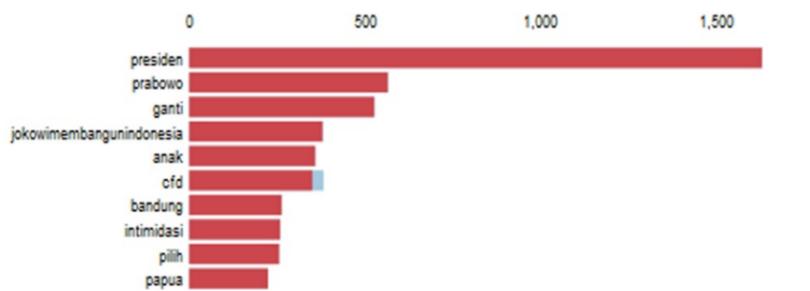

*Fig.5 Second topic of pro hashtags*

From figure 5, we summarize the second topic about one important incident, where it occurs intimidation towards a mother and her son which is done by some peoples wearing shirt labeled "*ganti presiden*" in a car free day event. Public using pro hashtag assume that the culprits behind this incident come from the supporter of Prabowo, the opponent of the incumbent. Public use pro hashtag also claim that they will not retaliate because it against moral or ethics value.

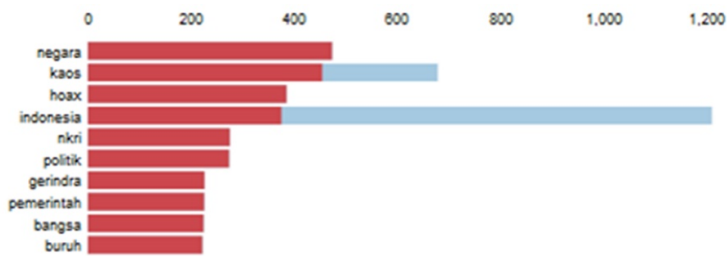

*Fig.6 Third topic of pro hashtags*

Moving towards the third topic as shown in figure 6, the topic is about the spreading of false rumour or hoax in the news. It is shown by terms *hoax* and *pemerintah*. This hoax is mainly about criminalization effort by the government to the several religious leaders in Indonesia. Some also talk about the Jokowi and his ex's picture on the internet, which is not true. These people using pro hashtag ask to those who involved in hoax activity to stop their action.

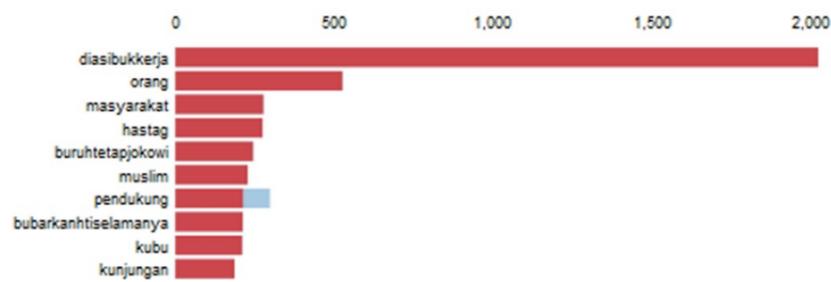

*Fig.7 Fourth topic of pro hashtags*

The fourth topic in figure 7 is about the support given by some labors to Jokowi. The term *buruhtetapjokowi* appears because of labors union who stand with political party named *Partai Demokrasi Indonesia Perjuangan* (PDIP). They claim that they will stand by PDIP because it has the same goal with labors in Indonesia. The aspiration of labors is admitted to be listened by them, hence it becomes their main reason to support on this side.

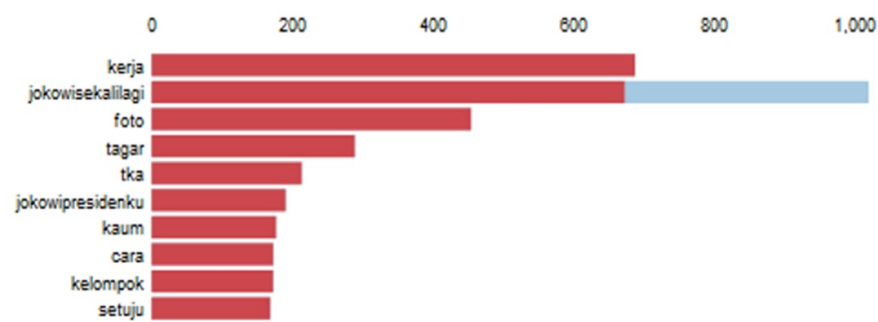

*Fig.8 Fifth topic of pro hashtags*

As the last top 5, we are able to see in figure 8 that the fifth topic is about foreign workers which is mainly shown by terms *kerja* and *tka*. This follows the executive order about the foreign workers that is released by Jokowi and eventually leads to pro and contra opinion among member of the society.

2) **Contra Hashtags**

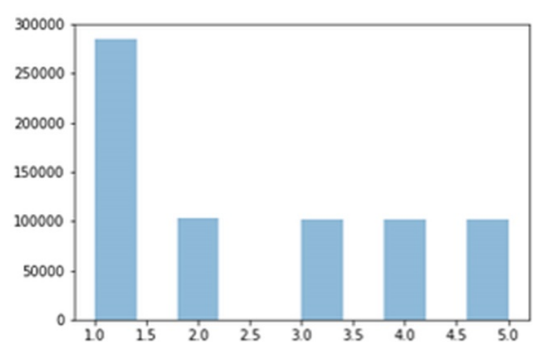

*Fig.9 Five major topics of contra hashtags*

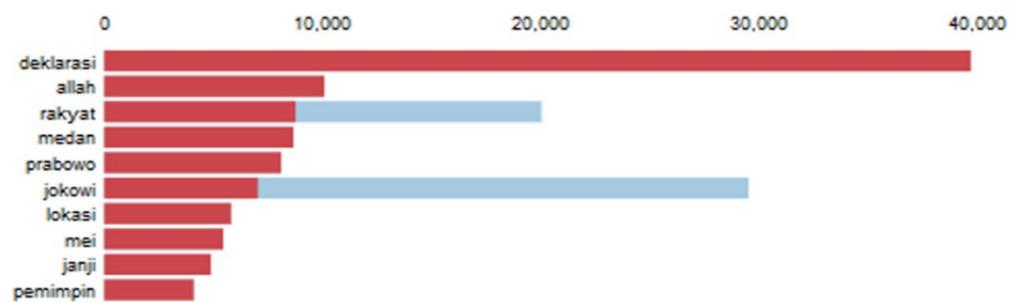
*Fig.10 First topic of contra hashtags*

In the figure 9, we can see the distribution of topics in the dataset generated by contra hashtag. Seeing the results in figure 10, we are able to see that the most dominant topic is about the big declaration of *#gantipresiden* that takes place in some cities in Indonesia. This can be seen from the term *deklarasi* which means declaration, as the most dominant term in the topic. We can also see term *Medan*, which is the name of the city, where the declaration event occurs. Medan becomes attentional because some local citizens in car free day event wear the shirt symbolize "*ganti presiden*", a supporter symbol against the incumbent. This probably occurs following the letter issued by local government, in order to discipline those people who, wear ganti presiden attribute in the car free day event.

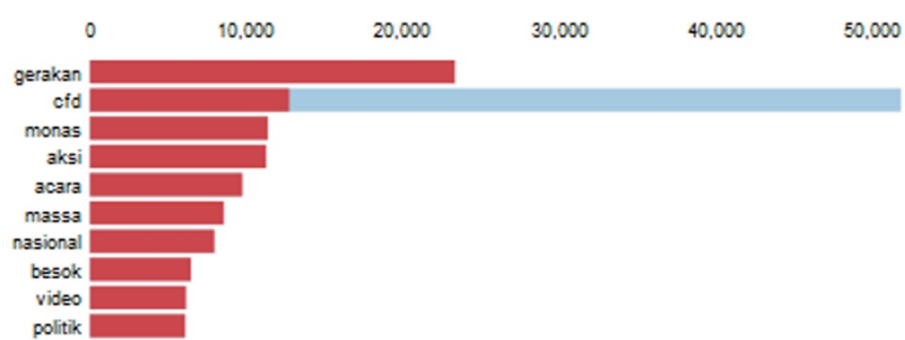
*Fig.11 Second topic of contra hashtags*

The second dominant topic in contra hashtag is about the demonstration which is done by people who join the march in National Monument. According to the news aired in the media, this demonstration will be held on May 6th, 2018, which is outside the range of our data collection scope. Hence, we can conclude that there are many people talking about the demonstration before the event had happened.

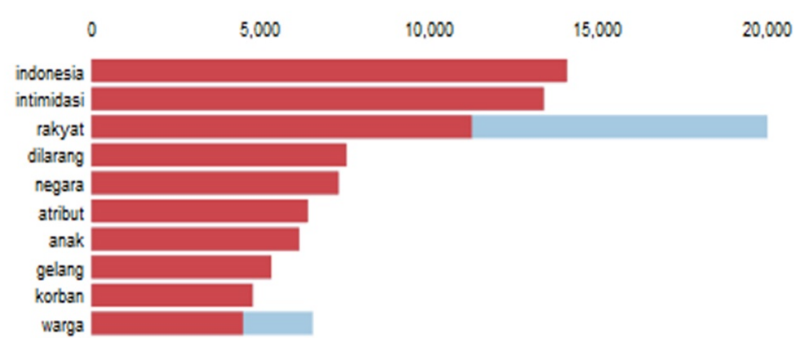
*Fig.12 Third topic of contra hashtags*

The third topic of this contra hashtags mainly tells us about the intimidation incident the same as the second biggest topic in pro hashtag. Both side of the opponents talk the same event using different tone. Mostly they reject the association or generalization of people who did the intimidation is Prabowo supporter, although those people wear Prabowo attributes when doing the intimidation.

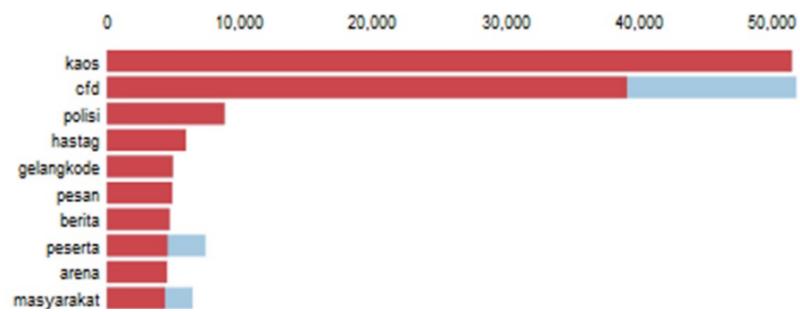

*Fig.13 Fourth topic of contra hashtags*

The fourth topic tells about the dismissal of people wearing the "*ganti presiden*" t-shirt by the police. We can see the most dominant term is *kaos* which means t-shirt, then followed by *cfd* which means car free day event. Next, we can see the term *polisi*, which mean police. In this topic, we check the dataset that people using contra hashtag talks about police who will dismiss the political event during that car free day event.

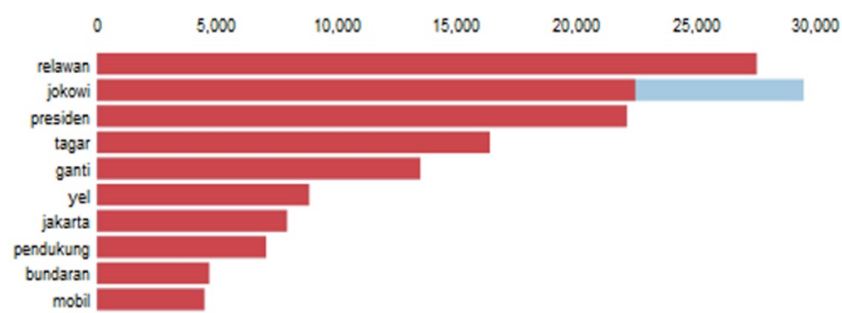

*Fig.14 Fifth topic of contra hashtags*

The last topic mainly talks about the volunteer's activity in the political event. Specifically, some people are talking about the volunteers of "*ganti presiden*" movement during the action in the *Bundaran HI* and water fountain in National Monument who are echoing their yells that it's being talked in Twitter communities.

**B. Social Network Analysis (SNA)**

From the observation, we get total of 118,144 actors and 544,075 relations based on pro and contra hashtags. Figure 15 shows the whole network visualization formed during 10 days data collection. The network consists of actors and edges from figure 15(a) look much denser when it compared to Figure 15(b). It shows that pro network has much more relations between actors. It also means that much more conversation are generated from this network.

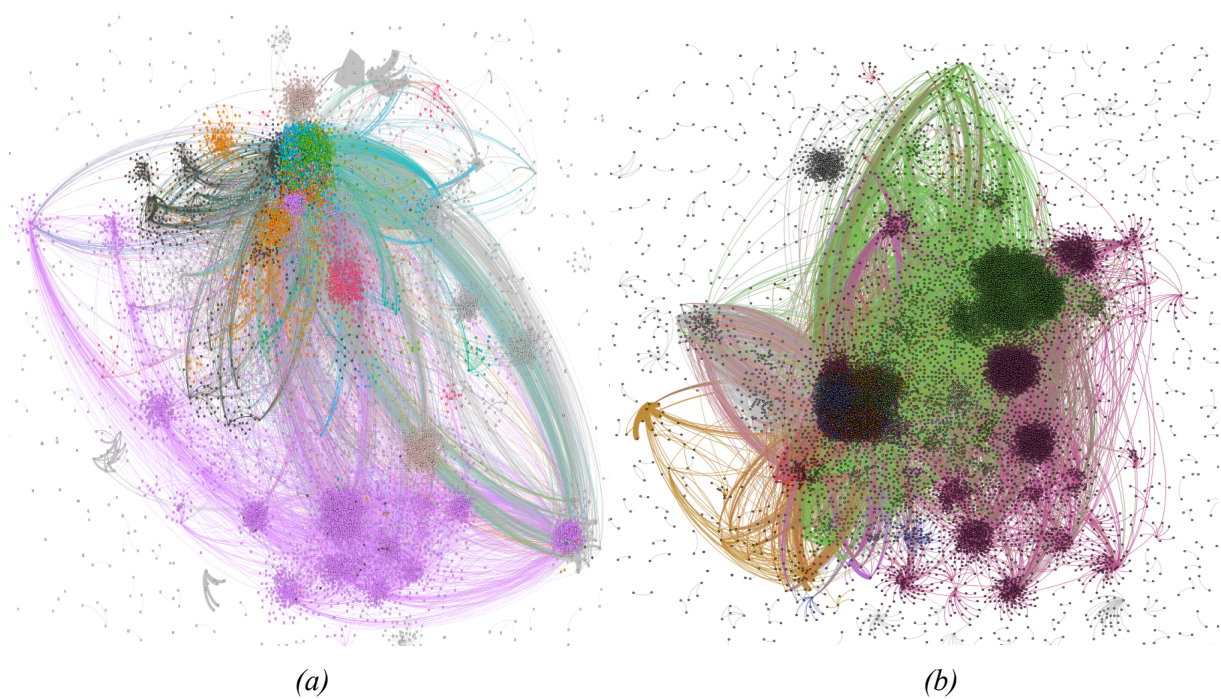

*(a)* *(b)*

*Fig.15. The social network of (a). Pro network, (*b)*. Contra network*

For the technical aspect, we employ the undirected network type that does not need to consider the direction of interactions, so the repeated parallel interactions are merged into one is chosen to visualize the network. After visualizing the social network, we elaborate from insights we get from the data from both sides. The network measurement metric of each opponent side shown at table 3.

*Table 3. Network Properties of Pro and Contra Networks*

| Network Properties | Contra | Pro |
|---|---|---|
| Tweets | 418,256 | 24,097 |
| Nodes | 88,353 | 29,791 |
| Edges | 462,224 | 81,851 |
| Average Degree | 6.79 | 7.509 |
| Diameter | 19 | 12 |
| Density | 0.219 | 0.607 |
| Modularity | 0.713 | 0.423 |

Number of the actors from contra network is almost 4 times higher than from pro network, it is shown from nodes measurement. Actor interactions from contra network is almost eighteen times higher than the pro one, it is shown from edges measurement. The average number of actor's interactions illustrated by average degree. It is shown that pro network is slightly higher than contra network. The higher average degree is, the faster information diffusion process from one actor to the rest of the network, hence we conclude that pro network is slightly faster to disseminate information. Network diameter shows the distance to reach the furthest actor in the network, in this case pro network is considerably shorter distance, means faster process. The modularity measurement shows how distinct the separation between group. Higher modularity value means the member of network are distinctly separated between group. Modularity values are in the range between 0 and 1. It shows that contra network is more distinct than pro network, that means actors in contra network exclusively belong to a particular group, while in pro network, the actor have tendency to be a member of different groups.

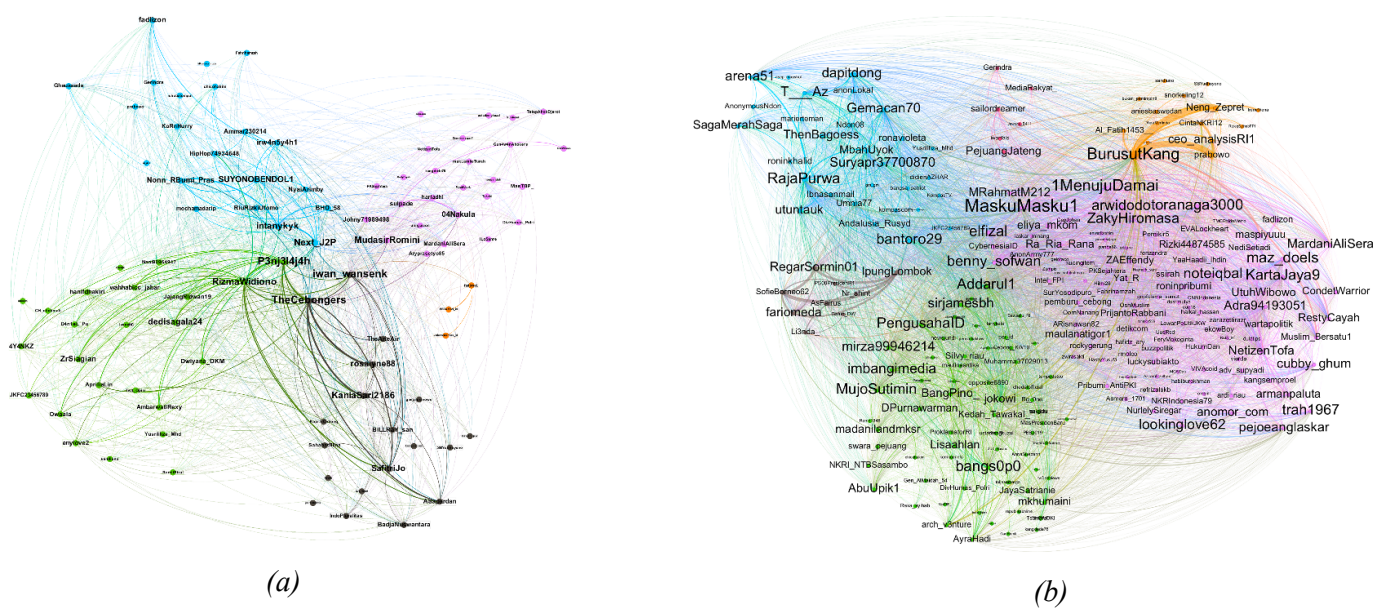

*(a)*      *(b)*

*Fig.16 The social network with actor identification of (a). Pro network, (b). Contra network*

The key player or key actor's measurement based on number of connections or acquaintance is called degree centrality metric. The highest degree centrality in pro network is an actor called @P3NJ3L4J4H, while in contra network is @RajaPurwa. These two actors have the biggest capacity to oversee the information dissemination since they are the greatest influence within their own network the network. Figure 16 shows the overall social network with actor identification.

**C. Dynamic Network Analysis (DNA)**

    **1) Node Evolution**

Node evolution shows the dynamic number of actors involved in conversation network during the observation period. Figure 17 shows the over time daily graph between pro and contra network. At some point on the 3rd day of observation, the number of actors in contra network suddenly arise sharply. By looking at the dataset, we see that is the April 29, 2018 (Friday), where the dominant topics is about Labor Day celebration, which will due in the coming few days. Labor Day topic comes from the efforts of the labor union movement to celebrate the economic and social achievements of worker. For two days straight, it was talking about it. Moreover, it happened on the weekend and most people have more free time, so that might be the reason there are quite high interactions among actors between from pro and contra network.

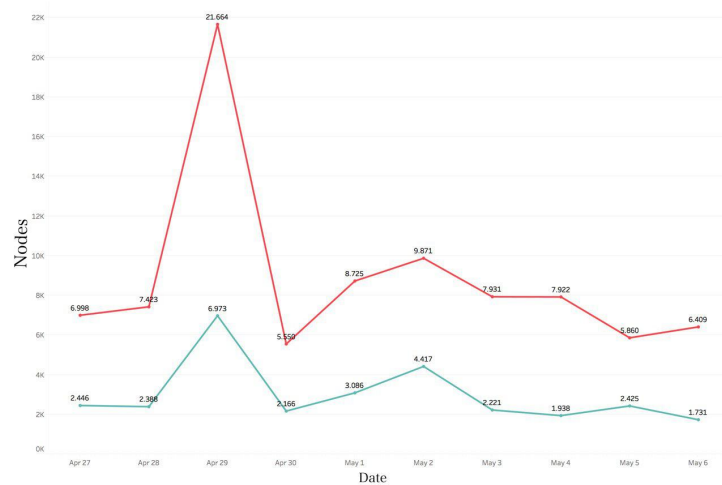

*Fig.17. Node dynamic of pro network (green) and contra network (red)*

2) **Edge Evolution**

Similar to node evolution, the edge evolution has peaked the number at the 3rd day of observations. This is reasonably normal, indicate the higher number of actors tends to create more conversations. Figure 18 shows the chart comparison during overall day of observation. Edge represent interactions which include tweets, replies, mentions and retweets. The conversations on the particular 3rd day is about Labor Day and Jokowi achievement for these past four years. They interact by asking and telling what improvement or degradation impacts of Jokowi action to Indonesia. It can be seen that the fluctuation of edges tends to occur more on the weekend.

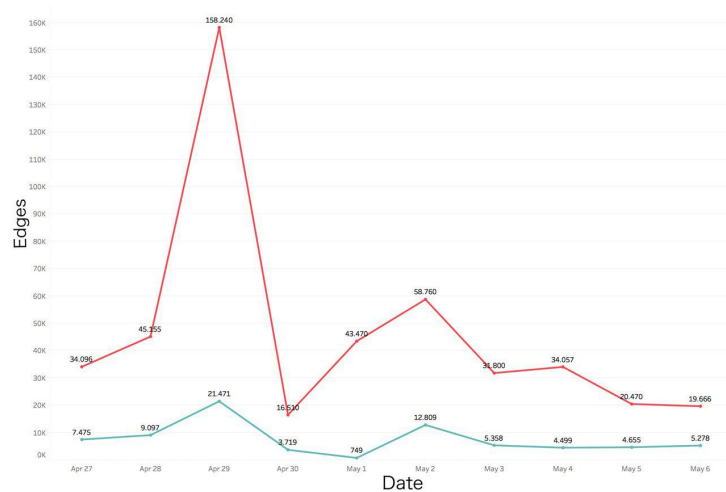

*Fig.18. Edge dynamic of pro network (green) and contra network (red)*

3) **Average Degree Evolution**

Figure 19 shows the daily dynamic of average degree measurement. The purpose of this metric is to see the average interactions of all actors in the network. For contra network, the 3rd day still hold the highest average degree value as followed by the 5th day and the 7th day. Whereas on pro network, the 2nd day has the highest value as followed by the 6th day and the 10th day. We detect the central nodes or key important actor with respect to spread the information and influencing others in their immediate neighborhood.

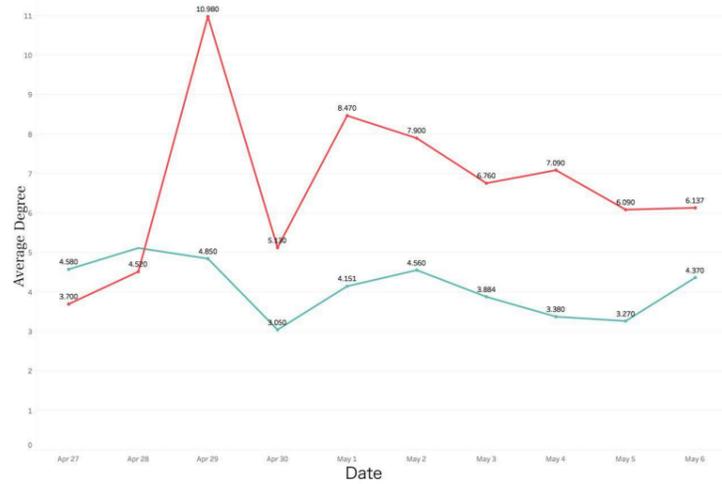

*Fig.19. Average degree dynamic of pro network (green) and contra network (red)*

### 4) Diameter Evolution

This analysis is used to determine the development of social network size characterized by network diameter during the observation period. Network diameter is the number of smallest steps that must be taken from a pair of furthest actors in the network. Smaller network diameter value indicates that information can travel faster. Thus, it will easily support faster information dissemination process. From the figure 20, the contra network diameters tend to more fluctuate compared to the pro network. The lowest network diameter value from pro network is 9 and contra network is 10. Related to the fluctuations experienced by both networks, the dynamic pattern of network diameter does not form a trend. However, it can be seen that pro network line graph tends to be more stable compared to contra network. It means that, the pro network easier to predict, it is shown by its capability to retain the network diameter value.

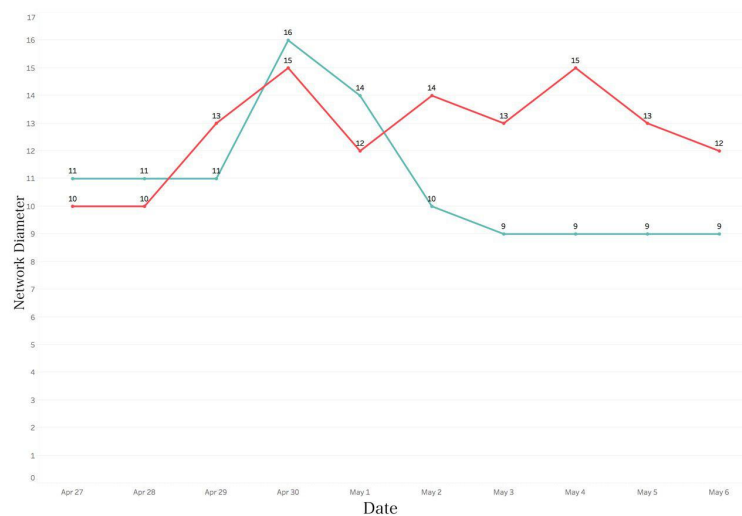

*Fig.20. Network diameter dynamic of pro network (green) and contra network (red)*

### 5) Modularity and Communities Evolution

We found number of communities varies according to daily interactions. Communities is determined by modularity measurement, which its main function is to detect the presence of the communities. Modularity itself measure the tendency of nodes grouping. The higher number of modularity means that the communities is distinct, or nodes are exclusively belonging to a community. Figure 21 shown that contra network has highest number of the communities on the 3$^{rd}$ day, while in figure 22 shown that highest modularity value from both networks are in the 4$^{th}$ day. We conclude that the distinct grouping on the 4$^{th}$ day reflect the stronger relationship inside the communities. It does not reflect to the number of the communities, where we have lower number of communities on the 4$^{th}$ day, but stronger relationship among its actors.

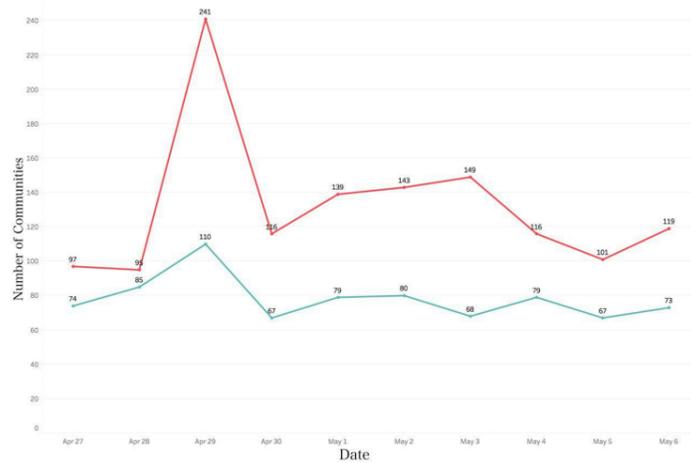

*Fig.21. Communities dynamic of pro network (green) and contra network (red)*

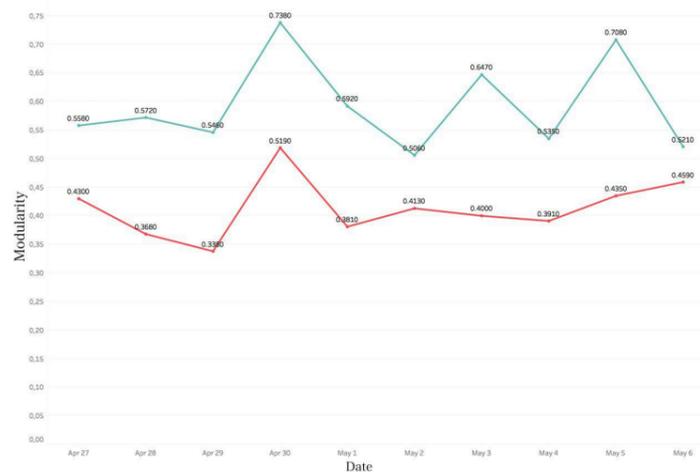

*Fig.22. Modularity dynamic of pro network (green) and contra network (red)*

6) **Density Evolution**

Network density measurement objective is to show how potential a network to become strong fully connected network. Network density calculate the ratio of number current edges to the maximum number of possible edges. The higher network density value means the denser or the closer relationship among actors in the network. Figure 22 shows that network density tends to decrease and showing the lower density value by time goes by. From the 6$^{th}$ day, network density reach value almost to zero, this means that either the connection is far less from the previous period or the increase number of node significantly, while the number of the edge is insignificantly increase.

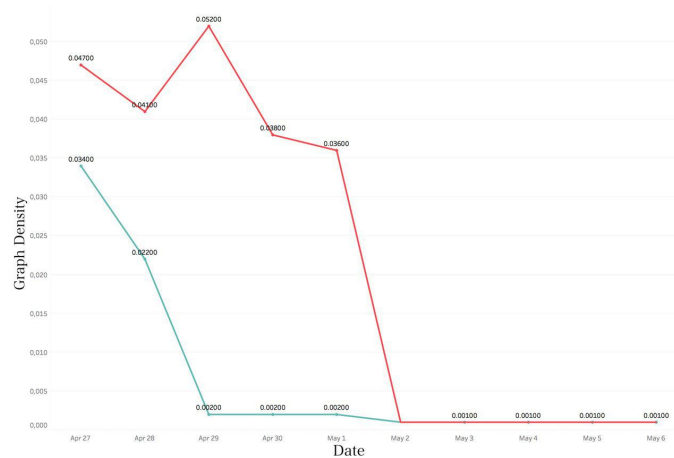

*Fig.22. Network density dynamic of pro network (green) and contra network (red)*

**D. Text Network Analysis (TNA)**

The pattern of relations between texts or terms that occur between pro and contra network can be seen in figure 23. The edges connecting the texts give us sense on what is the document context about. The edge thickness in figure 23 of both networks shows the strength edge's relation represent by weight. Thicker edge means the frequency of a pair of texts occur together in different phrases. The nodes thickness represents the frequency of particular texts shows up in the phrase or documents. The weight of both networks shown in table 4. To verify the result consistency, table 4(a) shows that *jokowi* and *president* text has 818 frequency, then we see from figure 23(a) that *jokowi* and *president* has the thickest edge.

Figure 23 also show that we are able to detect text network communities, shown by different color. In this sense, the communities formed in text network means the topics found in the particular network. For the example in figure 23(a), the first topic distinguished by purple color, it tells us about public support for Jokowi to run for the second period of presidency. In the contra network from figure 23(b), the text *ganti* and *presiden* have the thickest edge. Based on table 4(b), both term*s* have the highest weight among other relations, which is 12828. The contra network produces several topics, the first one is the purple color, which contain terms related to incident in the car free day event.

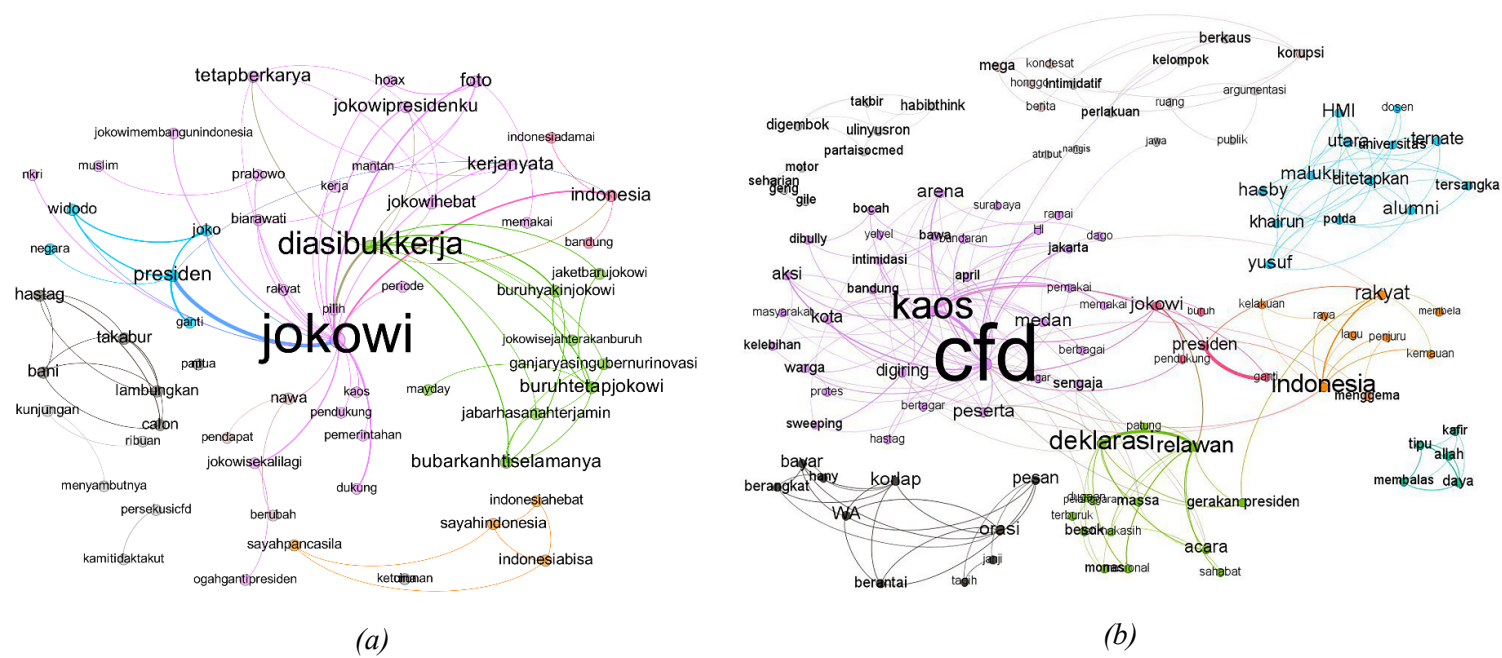

*(a)*                                                                                               *(b)*

*Fig.23 The text network of: (a) pro network, (b) contra network*

*Table 4 The top 14 relations of texts in: (a) pro network, (b) contra network*

(a)

| No. | Source | Target | Weight |
|---|---|---|---|
| 1 | jokowi | presiden | 818 |
| 2 | jokowi | diasibukkerja | 484 |
| 3 | ganti | Presiden | 389 |
| 4 | jokowi | indonesia | 351 |
| 5 | jokowi | dukung | 314 |
| 6 | jokowi | foto | 306 |
| 7 | joko | widodo | 277 |
| 8 | presiden | Joko | 276 |
| 9 | presiden | widodo | 272 |
| 10 | diasibukkerja | buruhyakinjokowi | 268 |
| 11 | jokowi | Jokowisekalilagi | 257 |
| 12 | diasibukkerja | Bubarkanhtiselamanya | 227 |
| 13 | jokowi | Jokowipresidenku | 214 |
| 14 | diasibukkerja | buruhtetapjokowi | 212 |

(b)

| No | Source | Target | Weight |
|---|---|---|---|
| 1 | ganti | presiden | 12828 |
| 2 | kaos | cfd | 11000 |
| 3 | deklarasi | relawan | 10536 |
| 4 | indonesia | rakyat | 4710 |
| 5 | cfd | aksi | 4546 |
| 6 | cfd | HI | 3958 |
| 7 | cfd | arena | 3936 |
| 8 | cfd | jakarta | 3891 |
| 9 | kaos | presiden | 3778 |
| 10 | tipu | daya | 3640 |
| 11 | kaos | memakai | 3601 |
| 12 | nasional | relawan | 3316 |
| 13 | jokowi | relawan | 3302 |
| 14 | jokowi | pendukung | 3261 |

# V. CONCLUSION

Opinion polarization is happening in Indonesia towards political event in 2019. We proof this by showing that one particular event appears and discussed on both opposite sides on different tone. The contradictory opinion on triggered by agenda to support or contra Jokowi. The four methodology TM, SNA, DNA, and TNA are used to reveal comprehensive understanding about polarization process. By understanding conversation content and topology, we are able to decipher the issues extracting from the content and the dynamic of conversation network. On topology network, by connect to the right people, we see how the issues disseminate and going viral. The interactions between actors in the social network go along dynamically. The network dynamics behavior gives the insights on how the network evolves when it is stimulated. It also helps us perceive the opinion evolution that goes in and out from each network where the contra side suddenly change to the pro side and vice versa.

This research gives us the understanding the whole process of virality and observe the interactions between actors timely. Social media forms a social network in some way similar to real-world networks as it represents the social behavior, shown by something that goes viral in the real world also becomes trending in social media. In addition, high traffic commonly happens during weekends and national celebration days comparing on weekdays. For instance, the action held by people who is contra Jokowi in some cities of Indonesia that happened on April 29th and the national celebration day of labor on May 1st create significant changing due to many important actors involved at that time.

To complete four methodologies above, we recommend to separate polarization directly using Sentiment Analysis methodology based on machine learning, as it gives the possibility to detect contradictory opinion while seemingly in pro or contra sides simultaneously. The second recommendation is the removal of preprocessing step to opponent classification; thus, it can be implemented on real time application or processing stream data. Furthermore, this will also help us capture the mechanism of false campaign, which campaign on right hashtag but carrying different message. Other suggestion is to involve more data and longer observation.